\documentclass[conference]{IEEEtran}
\IEEEoverridecommandlockouts
\usepackage{cite}
\usepackage{amsmath,amssymb,amsfonts}
\usepackage{algorithmic}
\usepackage{graphicx}
\usepackage{textcomp}
\def\BibTeX{{\rm B\kern-.05em{\sc i\kern-.025em b}\kern-.08em
    T\kern-.1667em\lower.7ex\hbox{E}\kern-.125emX}}
\usepackage{acronym} 
\usepackage[utf8]{inputenc}
\usepackage{listings}
    
\graphicspath{{figures/}{graphs/}}
\DeclareGraphicsExtensions{.png,.pdf,.jpeg}

\begin{document}


\acrodef{5gppp}[5G-PPP]{5G Infrastructure Public Private Partnership}
\acrodef{api}[API]{Application Programming Interface}
\acrodef{bbu}[BBU]{Baseband Unit}
\acrodef{bs}[BS]{Base Station}
\acrodef{capex}[CAPEX]{CAPital EXpenditure}
\acrodef{cord}[CORD]{Central Office Re-architected as a Datacenter}
\acrodef{cpri}[CPRI]{Common Public Radio Interface}
\acrodef{cran}[C-RAN]{Cloud Radio Access Network}
\acrodef{dc}[DC]{data center}%
\acrodef{epc}[EPC]{Evolved Packet Core}%
\acrodef{ewi}[EWI]{East-West Interface}%
\acrodef{gprs}[GPRS]{General Packet Radio Service}
\acrodef{gtp}[GTP]{GPRS Tunnelling Protocol}
\acrodef{gtpu}[GTP-U]{GPRS Tunnelling Protocol for User data}
\acrodef{hdn}[HDN]{Human-Defined Networking}
\acrodef{ide}[IDE]{Integrated Development Environment}
\acrodef{ietf}[IETF]{Internet Engineering Task Force}%
\acrodef{json}[JSON]{JavaScript Object Notation}
\acrodef{kpi}[KPI]{Key Performance Indicator}
\acrodef{lte}[LTE]{Long Term Evolution}
\acrodef{lldp}[LLDP]{Link-layer Discovery Protocol}
\acrodef{nbi}[NBI]{Northbound Interface}
\acrodef{nfv}[NFV]{Network Functions Virtualization}
\acrodef{odl}[ODL]{OpenDaylight}
\acrodef{of}[OF]{OpenFlow}
\acrodef{onf}[ONF]{Open Networking Foundation}
\acrodef{onos}[ONOS]{Open Network Operating System}
\acrodef{opex}[OPEX]{OPerating EXpenditure}
\acrodef{ovs}[OvS]{Open vSwitch}
\acrodef{ovsdb}[OVSDB]{Open vSwitch Database}
\acrodef{poc}[PoC]{Proof of Concept}
\acrodef{ran}[RAN]{Radio Access Network}
\acrodef{rat}[RAT]{Radio Access Technology}
\acrodef{rrh}[RRH]{Remote Radio Head}
\acrodef{sbi}[SBI]{Southbound Interface}
\acrodef{sdn}[SDN]{Software-Defined Networking}
\acrodef{xml}[XML]{eXtensible Markup Language}

\title{Towards Per-user Flexible Management in 5G}

\author{\IEEEauthorblockN{Aitor Zabala}
\IEEEauthorblockA{\textit{Research Department} \\
\textit{Telcaria Ideas S.L.}\\
Leganes, Spain \\
aitor.zabala@telcaria.com}
\thanks{This work is partially funded by the EU Commission in the frame of the Horizon 2020 projects SUPERFLUIDITY (grant no.  671566) and 5G-CORAL (grant no. 761586), and by grants from Comunidad de Madrid through Project TIGRE5-CM (S2013/ICE-2919).} 
\and
\IEEEauthorblockN{Elisa Rojas}
\IEEEauthorblockA{\textit{Departamento de Automatica} \\
\textit{Universidad de Alcalá}\\
Alcalá de Henares, Spain \\
elisa.rojas@uah.es}
\and
\IEEEauthorblockN{José María Roldan}
\IEEEauthorblockA{\textit{Research Department} \\
\textit{Telcaria Ideas S.L.}\\
Leganes, Spain \\
josemaria.roldan@telcaria.com}
\and
\IEEEauthorblockN{Luis Pulido}
\IEEEauthorblockA{\textit{Research Department} \\
\textit{Telcaria Ideas S.L.}\\
Leganes, Spain \\
luis.pulido@telcaria.com}
}

\maketitle
\begin{abstract}
Flexible management is one of the key components of next-generation 5G networks. Currently, many approaches focus on network functionality (services) and translate it afterward into end-user requirements, which slightly constrains the flexibility for both management and end users. Furthermore, moving the intelligence of the network towards the edge (i.e. the users) has already proven its benefits, such as computational offloading, lower latency and higher bandwidth utilization. In this article, we try to move management as close as possible to final users, providing per-user flexibility and unique user to service paths, enabling custom paths adapted for each user requirements instead of users adapting to service requirements. To validate our ideas, we work on two different use cases, implemented as proof-of-concepts in the ONOS platform. From the results obtained we conclude that there is still work to be done regarding the integration of SDN in the radio access and evolved packet core functions to provide the desired flexibility.
\end{abstract}


\section{Introduction}
\label{sec:intro}

On the road towards the fifth-generation mobile network (5G)~\cite{gupta:5g}, different challenges have been defined. The \ac{5gppp} lists various \acp{kpi} that should be accomplished, where particularly network management \ac{opex} is expected to be reduced by at least 20\% compared to today. 

One concept that has emerged recently together with 5G is the Mobile Edge Computing (MEC)~\cite{etsi:mec}, recently rebranded as Multi-Access Edge Computing, which conveys part of the network intelligence to the edge of it. The MEC framework is envisioned to leverage \ac{sdn}~\cite{kreutz:sdn} and \ac{nfv}~\cite{etsi:mano} technologies to enhance network management. 
Similarly, the \ac{cord} initiative~\cite{peterson:cord} has a branch focused on mobile networks (M-CORD) expected to procure proximity	to end users as well. 

Cloud technologies and flexible service management in \ac{ran} are key towards 5G~\cite{rost:cloud5g}. Accordingly, the network edge becomes a \ac{cran}~\cite{cran}, where the traditional \ac{bs} is cracked into two pieces: the \ac{rrh} (\textit{dummy} radio hardware) and the \ac{bbu} (which processes the baseband signal as part of the cloud intelligence)~\cite{i:cran}. This separation implies establishing a high-performance network in between: the fronthaul, whose traffic is generally transported by the \ac{cpri}~\cite{oliva:cpri}. 

Although the benefits of extending the cloud to the edge~\cite{montero:cloud-edge} seem apparent, the open question is how close can network management be from final users. 

In this article, we analyze how to leverage \ac{sdn} to manage per-user connectivity at the fronthaul in 5G and present a \ac{poc} implemented in \ac{onos}~\cite{onos-paper} (following the principles of \ac{cord}). Section~\ref{sec:approach} is devoted to the analysis and definition of the approach, Section~\ref{sec:eval} describes the implementation, and finally Section~\ref{sec:conclusion} examines conclusions and future work.



\section{A superfluid approach}
\label{sec:approach}

\begin{figure*}[htbp]
\centerline{\includegraphics[scale=0.45]{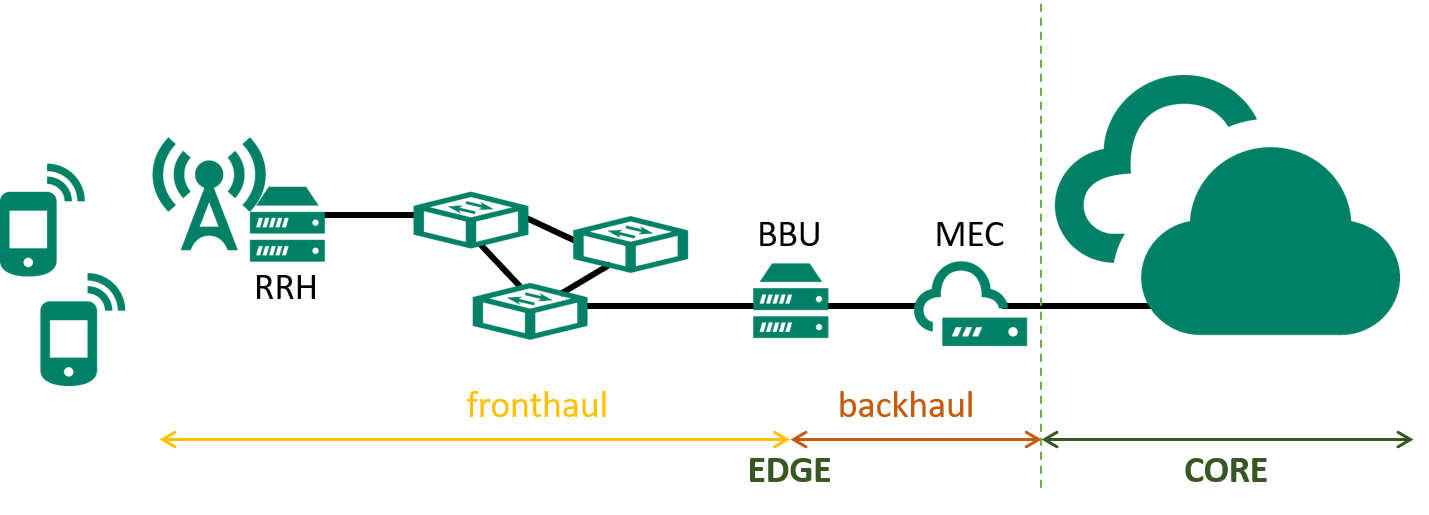}}
\caption{5G network architecture in Superfluidity}
\label{fig:5g}
\end{figure*}

The \textit{superfluid} network concept originates from the architecture for 5G defined in the Superfluidity project~\cite{bianchi:superfluidity}. One of its cornerstones is flexible network management. Ideally focused on supporting per-user granularity, services could be deployed at the core or at the edge (particularly, in the MEC) following end user requirements at each time. To go one step further, Superfluidity attempts to identify specific user profiles and bring them access accordingly. 

To accomplish it, first we decided to design a management framework focused on users and their associated services (instead of services and the users who employ them). Second, we try to merge both fronthaul and backhaul management, to avoid defining a barrier in between. Third, we focus on two use cases: bandwidth control and traffic paths based on users. Finally, we analyse the current protocols in play for those use cases, mainly \ac{gtpu} and \ac{cpri}. The following sections detail these aspects in further detail.

\subsection{Per-user flexible management}
So far, 5G architectures first focus of the services deployed and, later on, on how they affect end users. The principle we want to accomplish is the other way round: individual people connecting to the mobile network anytime, anywhere, while the network adapts to their requirements, following the principles of \ac{hdn}~\cite{rojas:hdn}. Hence questions emerge: Would that user have all the services at all places? Could they configure what they want at any time (adjusting billing accordingly)? What if they want to hire a service at that right moment without needing to call their provider?

In line with the above questions some examples arise, such as, paying for a new service at user demand, one idea would be having a dynamic user captive portal from which end users could configure their services.

\subsection{Backhauling the fronthaul}
To accomplish per-user flexible management, the immediate question is where should the management framework be located in the network. According to the current deployment of Superfluidity (see Fig.~\ref{fig:5g}), which follows the standard 5G network architecture, the edge network is divided into two pieces: fronthaul and backhaul. More specifically, the fronthaul is the closest path to the user and, ideally, per-user management should start from it. Currently, the fronthaul is just thought as a --non-cloudified-- transport network. However, we believe it should also be part of the cloud and the associated management framework. The barrier between fronthaul and backhaul should be blurred, as stated in~\cite{oliva:xhaul}.

\subsection{Two use cases: per-user bandwidth control and traffic diversion}
We envision two main use cases:
\begin{enumerate}
  \item \textbf{Per-user bandwidth control:} By leveraging \ac{sdn}, we could control the fronthaul transport network, not only dynamically instantiating paths between the \acp{rrh} and \acp{bbu}, but also defining the characteristics of those paths, such as bandwidth.
  \item \textbf{Per-user traffic diversion:} Diverting specific user traffic could be useful for different scenarios, such as the user captive portal we mentioned in the example above, or even for security reasons, e.g. the user's mobile is temporarily hacked and we want to drop malicious traffic as close as possible to the user.
\end{enumerate}

\subsection{The protocols in play: \ac{gtpu} and \ac{cpri}}
In traditional \ac{lte} architectures, the data plane transport protocol on the core network is based on \ac{gtp}, which is the protocol carrying \ac{gprs} packets. Inside \ac{gtp} we can find three other sub-protocols: i) GTP-C, mainly used for signalling, session activation and QoS provisioning for users, ii) GTP-U, focused in transporting user data through the data plane, and iii) GTP', focused on billing.
 
Alternative approaches have been designed in order to remove the dependency with GTP tunnels. M-CORD~\cite{mcord} redefines the radio access, by levering \ac{sdn} and \ac{nfv} technologies, to deploy flexible networks, capable of acting as cloud platforms to deploy services. M-CORD features: i) Programmable \ac{ran} (SD-RAN), ii) Disaggregated and virtualized \ac{epc}, iii) MEC, and iv) End-to-end slicing from \ac{ran} to \ac{epc}. On the fronthaul ONOS controls the virtualized \ac{bbu} instances deployed to serve their \acp{rrh}. With such control over the data plane and disaggregated EPC functionalities, M-CORD can progressively remove GTP connection tunnels in favor of OpenFlow tunnels. Migrating to a connectionless LTE. 
 
 On the other hand, nowadays \ac{cpri} is the preferred option to deploy the fronthaul network. Introducing this interface in the management framework would allow a new range of possibilities of control closer to end users.

\section{Implementation and Evaluation}
\label{sec:eval}

To demonstrate the feasibility of the two use cases described in the previous section, we implemented a proof-of-concept of each of them levaraging the \ac{onos} platform as \ac{sdn} controller (more specifically \ac{onos} version 1.10.4). The network was both virtualized with the Mininet~\cite{mininet} platform, using \ac{ovs}~\cite{ovs}, and also tested with the hardware \ac{sdn} switch Pica8~\cite{pica8} (P-3297 model), which is \ac{ovs}-based as well.

\subsection{Per-user bandwidth control}
The fronthaul could be simplified as a transport network in between of two final nodes: the \ac{rrh} and the \ac{bbu} (as represented in Fig.~\ref{fig:5g}). At the same time, \ac{onos} provides the \textit{Intent Framework}~\cite{onos-intent}, which automatically deploys shortest paths between pairs of end hosts, for example. An \textit{intent} is a way of expressing \textit{what} functionality we want in the network, instead of focusing on \textit{how} to do it. Accordingly, thanks to \ac{onos}, we can easily build a minimum latency path between the \ac{rrh} and the \ac{bbu} (without worrying about the underlying network), but that path will have the same characteristics for all end users served. Thus, our objective was to implement some type of differentiation in the paths between the \ac{rrh} and \ac{bbu} nodes.

\begin{figure}[htbp]
\centerline{\includegraphics[scale=0.8]{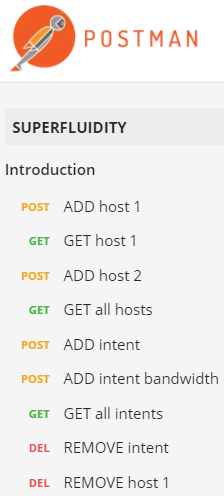}}
\caption{REST API definition in Postman of the \textit{intent bandwidth} app}
\label{fig:postman}
\end{figure}

To start with, we decided to enhance the current intent framework in \ac{onos} by developing a \textit{bandwidth-based intent}, i.e. shortest path with an additional parameter for QoS: bandwidth control. More specifically, we implemented an \ac{onos} app that introduces \textit{bandwidth-based intent} and a REST API for control, as depicted in Fig.~\ref{fig:postman}. The REST API permits dynamic instantiation of host scenarios from scratch and intent deployment afterwards. 

For example, initially the network manager will be able to define the hosts that he is willing to connect. These hosts can be of type \ac{rrh} (\texttt{ADD host 1}) or \ac{bbu} (\texttt{ADD host 2}). An example of how an \ac{rrh} node would be added is shown in Listing~\ref{list1}.

Once hosts and their roles are settled, we can dynamically create the shortest paths with delimited bandwidth (\texttt{ADD intent bandwidth}) between pairs of RRH and BBU hosts, as shown in Listing~\ref{list2}. This new type of intent leverages the already existing \texttt{HostToHost} intent in \ac{onos}. Note the simplicity of the command, which only requires 3 parameters to set up routes: the \ac{rrh}, the \ac{bbu}, and the bandwidth (optional parameter). 

\begin{lstlisting}[frame=single, basicstyle=\small\ttfamily, columns=flexible, caption={Usage sample of the \texttt{ADD host} command},captionpos=b, label=list1]
curl --request POST \
  --url http://IP:p/superfluidity/edge/host/ \
  --header 'accept: application/json' \
  --header 'authorization: Basic <...>=' \
  --header 'content-type: application/json' \
  --data '{
  "device":"of:0000000000000001",
  "port":"1",
  "mac":"00:30:18:c9:ef:cd",
  "vlan":"-1",
  "ips":["172.16.1.1"],
  "type":"RRH"
}'
\end{lstlisting}

\begin{lstlisting}[frame=single, basicstyle=\small\ttfamily, columns=flexible, caption={Usage sample of the \texttt{ADD intent bandwidth} command},captionpos=b, label=list2]
curl --request POST \
  --url http://IP:p/superfluidity/edge/intent/ \
  --header 'accept: application/json' \
  --header 'authorization: Basic <...>=' \
  --header 'content-type: application/json' \
  --data '{
	"one":"00:00:00:00:01:01/None",
	"two":"00:00:00:00:02:01/None",
	"bandwidth":100
}'
\end{lstlisting}

The implemented app is composed of the following modules:
\begin{itemize}
  \item \texttt{EdgeHost}: Edge host model.
  \item \texttt{EdgeIntent}: Edge intent model.
  \item \texttt{EdgeHostWebResource}: It implements the REST API for the hosts (add, delete, etc.).
  \item \texttt{EdgeIntentWebResource}: It implements the REST API for the intents (add, delete, etc.).
  \item \texttt{EdgePacketProcessor}: It is in charge of detecting traffic between one \ac{rrh} and a \ac{bbu}.
  \item \texttt{InternalHostProvider}: It extends the \texttt{AbstractProvider} class to be able to add, modify and delete hosts in \ac{onos}.
  \item \texttt{EdgeComponent}: Main class. It implements the \texttt{Activate} and \texttt{Deactivate} methods and it contains the rest of the implementation.
  \item \texttt{EdgeService}: Class that defines the service to be implemented in EdgeComponent.
\end{itemize}

\begin{figure}[htbp]
\centerline{\includegraphics[scale=0.35]{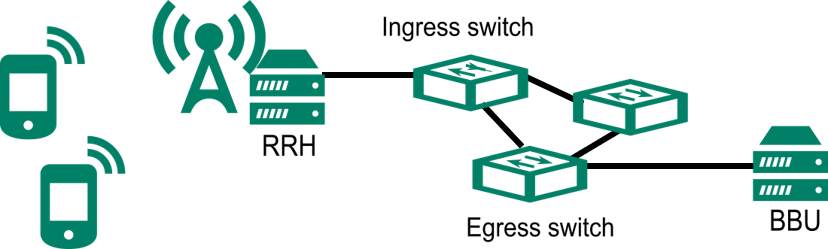}}
\caption{Representation of the ingress and egress switches in the fronthaul architecture (simplified view, as usually more switches will shape the fronthaul)}
\label{fig:rrh-bbu-switches}
\end{figure}

To install the different paths, the app connects to the network switches through the \ac{ovsdb} protocol~\cite{ovsdb}. The app installs a QoS entry and an associated queue in the ingress and egress switches (as represented in Fig.~\ref{fig:rrh-bbu-switches}, which guarantees the requested bandwidth limitation when the route is created after the REST API call. Once the QoS entry and the queues are installed, the \texttt{HostToHost} intent is installed by specifying the two hosts we want to connect and the queue ID we want to link to the intent. Currently, \texttt{HostToHost} intents can only be associated to a single queue ID, so the queue ID of the ingress and egress switches are set to the same value.

\begin{figure}[htbp]
\centerline{\includegraphics[scale=0.5]{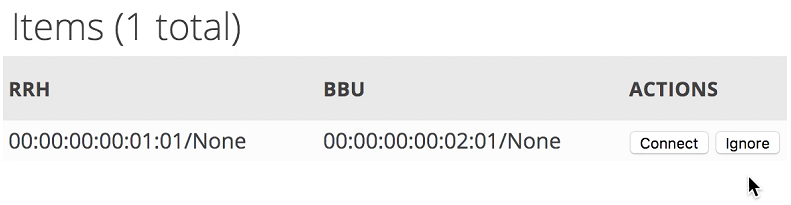}}
\caption{ONOS GUI detecting a connection between RRH and BBU}
\label{fig:rrh-bbu}
\end{figure}

Additionally, the app implements an extension of the \ac{onos} GUI that is able to alert the network manager when traffic is detected between a \ac{rrh} and a \ac{bbu}, which potentially would require a path to connect them both. The extension of the GUI is shown in Fig.~\ref{fig:rrh-bbu}. To integrate this frontend web with the \ac{onos} platform, the following modules were developed:
\begin{itemize}
  \item \texttt{EdgeUiTableComponent}: Integration of the frontend web as a table component in \ac{onos}. 
  \item \texttt{EdgeUiTableMessageHandler}: Implementation of the app extended GUI.
\end{itemize}

\begin{figure}[htbp]
\centerline{\includegraphics[scale=0.65]{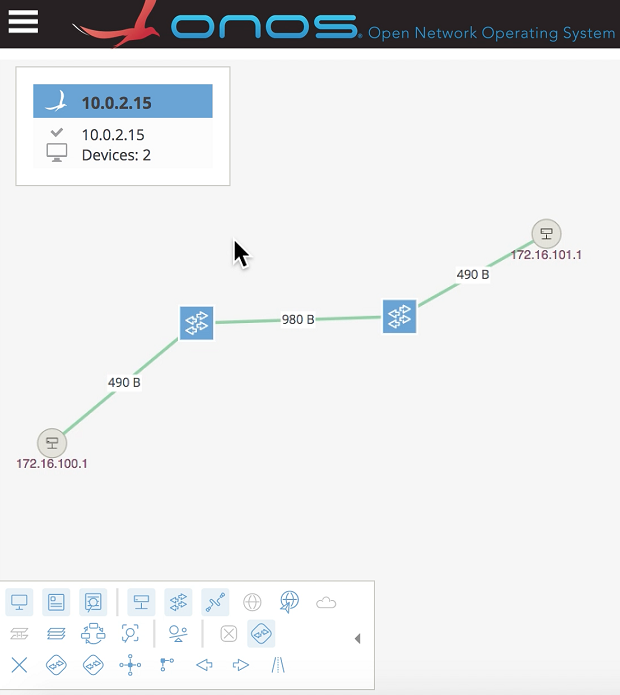}}
\caption{ONOS GUI graphically showing the traffic between RRH and BBU}
\label{fig:rrh-bbu-running}
\end{figure}

Finally, we tested the implementation in a Mininet \ac{ovs}-based network. The graphical representation of the traffic is depicted in Fig.~\ref{fig:rrh-bbu-running}. However, although routes were installed correctly at every moment, after testing with \texttt{ping} and \texttt{iperf}, we realized that bandwidth limitations was not being granted. After studying the tests carefully, we discovered that \ac{onos} was performing correctly and the problem was related with how Mininet implements links between the \acp{ovs}.

Therefore, we repeated the tests on a network built with the P-3297 switch from Pica8. Pica8 internally implements \acp{ovs} as well, in a customized way. This time the problem was a different one, which is that \ac{onos} raises a \texttt{code=BAD\_QUEUE} in the logs, and flows are not installed, kept in the \texttt{PENDING\_ADD} status. \\
After analyzing the Pica8 documentation, we found out that the port QoS is Pica8 specific and should be one of the following: 
\texttt{type=PRONTO\_STRICT} or \texttt{type=PRONTO\_WEIGHTED\_ROUND\_ROBIN}. \\
Currently, neither of them is implemented by \ac{onos}, according to their code in \texttt{QosDescription.java}.

Up to this point, we stopped the implementation, as we considered some development effort needs to be done from the communities of Mininet, Pica8 and \ac{onos}. A relatively fair solution would be updating the \texttt{QosDescription.java} code, but it implies both people from Pica8 and \ac{onos} to work together.
\subsection{Per-user traffic diversion}
The second use case implies processing per-user granular flows by analyzing the traffic inside \ac{gtpu} tunnels. The main objective is to divert specific user traffic before it reaches the cloud, i.e. the service location. As we mentioned in the introduction, several reasons might require this diversion, such as: securing the network from potential attacks (diverting traffic to honeypots) or directing a user to a captive portal for service hiring.

To implement this use case proof-of-concept, the first requirement is the support of \ac{gtpu} tunneling in \ac{ovs}, which is the switch we are using for our tests, and also the most supported along the \ac{sdn} community. Currently, \ac{ovs} needs a patch to support it~\cite{gtpu-ovs}.

\begin{figure}[htbp]
\centerline{\includegraphics[scale=0.35]{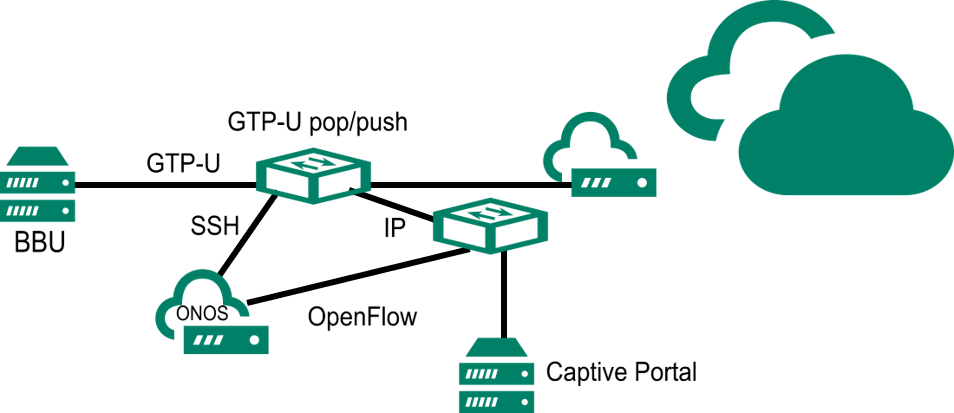}}
\caption{Representation of user captive portal in the architecture}
\label{fig:rrh-bbu-portal}
\end{figure}

In this case, the app we developed leverages the previous feature to encapsulate and decapsulate \ac{gtpu} traffic. More specifically, a switch associated to the \ac{bbu} is in charge of this task, as shown in Fig.~\ref{fig:rrh-bbu-portal}. Thanks to it, we can distinguish the owner (end user) of the traffic and act accordingly. \\
To follow this behaviour, we install the flows in the switch via an SSH tunnel. The reason behind this is that OpenFlow does not support the installation of these flows. Thus, one alternative design would be extending the OpenFlow protocol so that it supports the use and inspection of \ac{gtpu} frames.

A very simplistic scenario would be the following: A user hired voice, but not a data service with the operator. However, she decides to start using some data services, so the first time she uses \ac{gtpu}, her traffic is diverted to a user captive portal where she can directly hire new services by adding her billing information (as depicted in Fig.~\ref{fig:rrh-bbu-portal}). This portal would communicate with the \ac{sdn} platform (e.g. via REST API), which would install the corresponding paths afterwards. \\ 
This minimizes the effort of the network manager, who usually activates services manually, and it also gives added control and enhances service flexibility to end users, who are capable of deciding which services they want to use anywhere at anytime.

\section{Conclusions and Future Work}
\label{sec:conclusion}

Along the article, we have described the need to move network management towards the edge. The main objetive is to provide management flexibility with focus on end users, following the principles of \ac{hdn}.

To prove our ideas, we have described two use cases: (1) per-user bandwidth control and (2) per-user traffic diversion, and we have implemented a proof-of-concept of each of them as an app in the \ac{onos} \ac{sdn} platform. 

The first use case proves the feasibility of the approach, reducing management time (i.e. a command with just 3 parameters installs shortest path with specific QoS) while adding flexibility at the same time (nodes are easily registered via a REST API and tracked via a graphical interface). However, we could not test the QoS as Mininet does not implement bandwidth limitation, and Pica8 requires specific queues, which are not currently supported in \ac{onos}. As future work, we consider the Pica8 and \ac{onos} community should collaborate for integration of this feature.

The second use case adds flexible management both for the services provider and end users, who are capable of hiring their services at request. However, we had to install a patch in \ac{ovs} to support \ac{gtpu} tunnelling, and flows are installed via SSH, instead of OpenFlow, which does not support it. As future work, we believe extending OpenFlow to support \ac{gtpu} should be considered, or at least new Southbound Interface protocols under research should have it in mind.






\bibliographystyle{IEEEtran}
\bibliography{sf-scene} 


\end{document}